\documentclass{article}
\usepackage{graphicx}
\usepackage{url}

\usepackage{amssymb}
\usepackage{amsmath}
\usepackage{latexsym}

\newcommand{\figlab}[1]{\label{fig:#1}}

\newcommand{\figref}[1]{\ref{fig:#1}}

{\catcode`\@=11
\gdef\setft#1#2#3{%
\def\@oddfoot{
{\setbox0=\hbox{#1}
\setbox1=\hbox{#3}
\ifdim\wd0>\wd1
\dimen0=\wd0
\box0\hfil#2\hfil\hbox to\dimen0{\hfil\hfil\box1}
\else \dimen0=\wd1
\hbox to\dimen0{\box0\hfil }\hfil#2\hfil\box1 \fi
}}} }

\def\complaint#1{}
\def\withcomplaints{
\newcounter{mycomplaints}
\def\complaint##1{\refstepcounter{mycomplaints}%
\ifhmode%
\unskip%
{\dimen1=\baselineskip \divide\dimen1 by 2 %
\raise\dimen1\llap{\tiny -\themycomplaints-}}\fi%
\marginpar{\tiny [\themycomplaints]: ##1}}%
}

\title{Computational Geometry Column 44}
\author{%
Joseph O'Rourke\thanks{
Dept. of Computer Science, Smith Col\-lege, North\-ampton, 
MA 01063, USA.
orourke@\allowbreak cs.\allowbreak smith.\allowbreak edu.
Supported by NSF Distinguished Teaching Scholar Grant DUE-0123154.
}
}
\date{}
\begin{document}
\maketitle

\begin{abstract}
The open problem of whether or not every pair of equal-area polygons
has a hinged dissection is discussed.
\end{abstract}

A \emph{dissection} of one polygon $A$ to another $B$ is a partition of $A$
into a finite number of pieces that may be reassembled to form $B$.
$A$ and $B$ are then said to be \emph{equidecomposable}.
A staple of puzzle collections for centuries, the field has been given new
life through Frederickson's 1997 monograph~\cite{f-dpf-97}.
He quickly followed that with a second book~\cite{f-hdst-02} on \emph{hinged dissections},
the most famous of which is the elegant 1902 Dudeney-McElroy hinged
dissection between a square and an equilateral triangle illustrated in
Fig.~\figref{sq.tri.hinged}.
\begin{figure}[htbp]
\centering
\includegraphics[width=\linewidth]{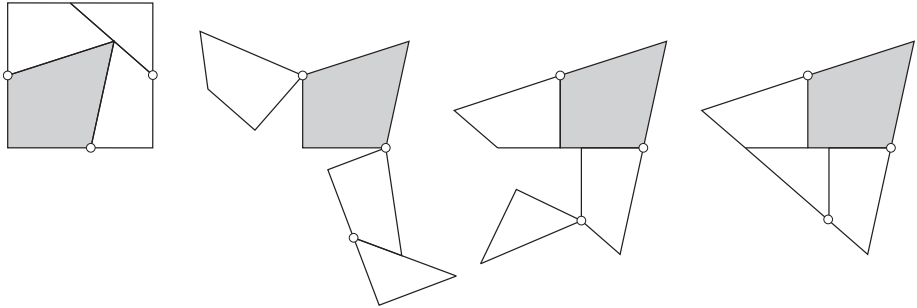}
\caption{A four-piece hinged dissection between a square and an equilateral triangle.}
\figlab{sq.tri.hinged}
\end{figure}

It has been known since the early 19th century that any two polygons of equal
area are equidecomposable.\footnote{
  It has been known since the early 20th century that not all equal-volume
  polyhedra are equidecomposable.}
One method~\cite[p.~221]{f-dpf-97}
triangulates polygon $A$, dissects each triangle to a rectangle,
converts each rectangle to one of fixed length, and stacks all the rectangles.
Applying the same process to $B$ and overlaying the stacked rectangles yields
a mutual dissection.
No comparable method is known for hinged dissections:

\begin{center}
\fbox{
\begin{minipage}[h]{0.75\linewidth}
{\sc question:} Does every pair of equal-area polygons have
a hinged dissection?
\end{minipage}
}
\end{center}

\noindent
The usual interpretation of this problem ignores possible intersections between
the pieces as they hinge, 
following what Frederickson calls the ``wobbly-hinged'' model.  
Removing this freedom is perhaps not difficult, but seems not to
be settled in the literature.

Here I report on two advances on this problem.
The first is a proof that any two \emph{polyominoes} of the same area
have a hinged dissection~\cite{ddeff-hdpp-03}.
A polyomino is a polygon formed by joining unit squares at their
edges~\cite{g-p-65,k-p-97}. One composed of $n$ squares is called an \emph{$n$-omino}.
The authors establish a type of universal dissection by proving that
a cycle of $2n$ right isosceles triangles, hinged at their base vertices,
can fold to any $n$-omino.
See Fig.~\figref{polyomino}.
Although the cycle may seem a special case, they prove that any hinged
graph structure may be converted into another dissection hinged
in a cycle.
\begin{figure}[htbp]
\centering
\includegraphics[width=\linewidth]{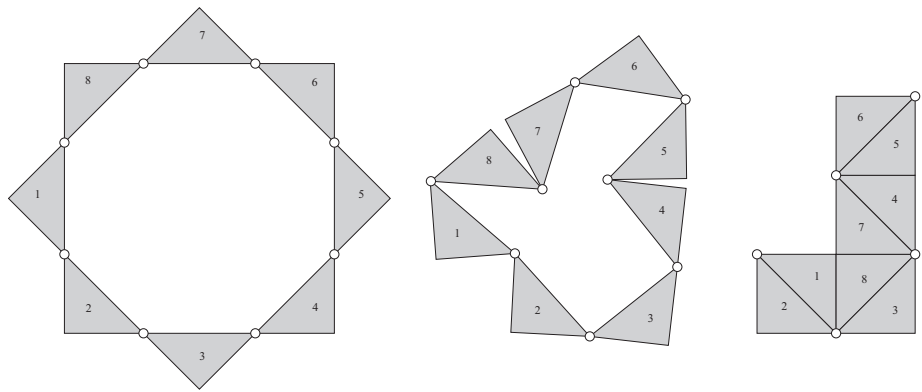}
\caption{Eight right isosceles triangles hinging to a tetramino.}
\figlab{polyomino}
\end{figure}
The polyomino result generalizes to hinged dissections 
of all edge-to-corresponding-edge gluings
of congruent copies of any polygon.
This leads, among other curiosities, 
to a $128$-piece hinged chain of isosceles right triangles
that folds to shapes representing
all the letters of the
alphabet and the ten digits~\cite{HingedAlphabet}.

The second advance is a proof by Eppstein~\cite{e-hkmd-01} that
any asymmetric polygon has a hinged dissection to its mirror image.
His dissection partitions the polygon into a chain of \emph{kites},
quadrilaterals with reflection symmetry across a diagonal.
With this partition in hand, it is easy to rotate the kites so all
their symmetry diagonals align.  The mirror image of
the unfolding process then folds the shape to its mirror image.
This procedure is illustrated for a simple example in Fig.~\figref{kites}.
\begin{figure}[htbp]
\centering
\includegraphics[width=\linewidth]{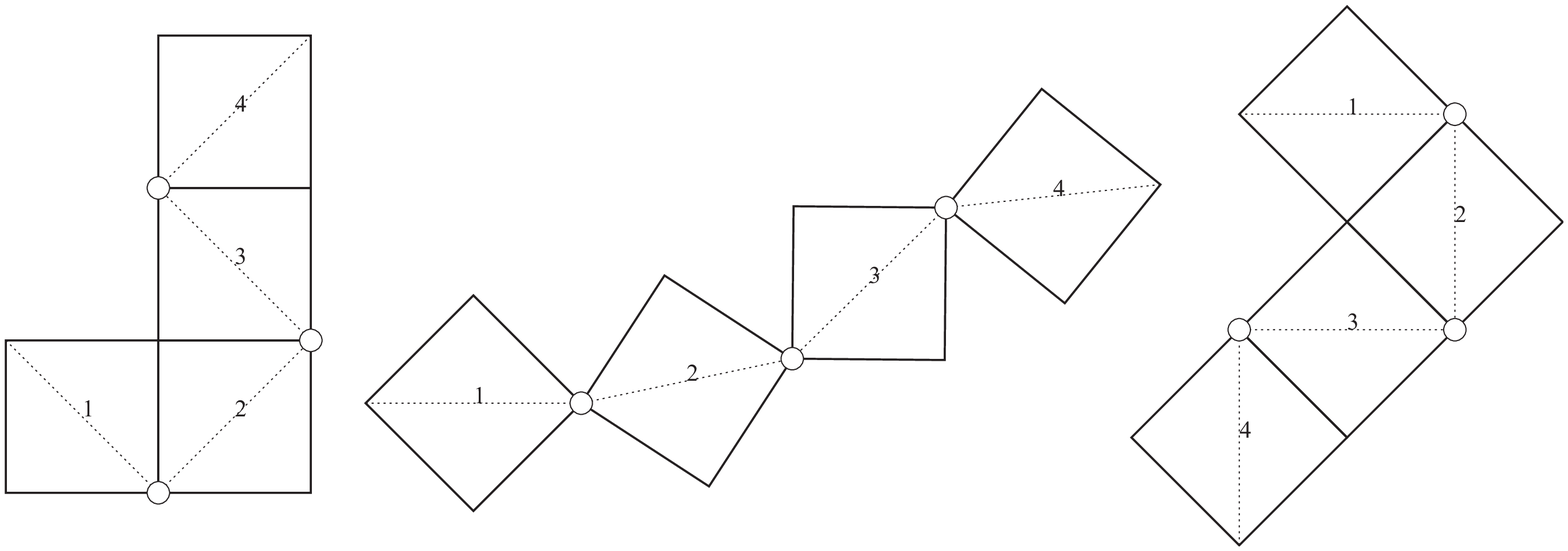}
\caption{A dissection into a chain of kites,
hinging to the mirror image.  Here the kites are all squares.}
\figlab{kites}
\end{figure}
The kite partition relies on circle packing~\cite{BerEpp-IJCGA-00},
and must be supplemented by a refined partition, and threading
around the boundary of a spanning tree to obtain a hinged chain.
This work leads to a possible route toward solving
the general problem for hinged dissections, for it reduces
it to $O(n)$ equal-area triangle hinge-dissection problems with particular constraints
on the hinges.

\small
\bibliographystyle{/home1/orourke/tex/alphasort}
\bibliography{44,/home1/orourke/bib/geom/geom}

\end{document}